\documentclass[10pt]{article}

\usepackage{graphics}
\usepackage{graphicx}

\usepackage[a4paper, left=35mm,right=35mm,top=34mm,bottom=34mm]{geometry}
\usepackage[utf8]{inputenc}
\usepackage[T1]{fontenc}
\usepackage[english]{babel}

\usepackage{enumerate}
\usepackage{graphicx}
\usepackage{hyperref}
\hypersetup{
    colorlinks=true,
    linkcolor=blue,
    filecolor=magenta,      
    urlcolor=cyan,
}
\usepackage{listings}
\usepackage{color}

\definecolor{dkgreen}{rgb}{0,0.6,0}
\definecolor{gray}{rgb}{0.5,0.5,0.5}
\definecolor{mauve}{rgb}{0.58,0,0.82}
\lstdefinelanguage{MRGC++}{%
  language=C++,
  morekeywords={T, U, MPI_Irecv, MPI_Isend, MPI_Allreduce, MPI_Waitall, Compute, Map, abs, max, Swap, MPI_Recv_init, MPI_Send_init, MPI_Startall, Copy, Init, InitRecv, InitSend, InitAllReduce, Send, Recv, AllReduce, Finalize, InitSnapshot, Snapshot, SwitchAsync, SnapReduce, MPI_Test, MPI_Start}
}
\usepackage{mathtools,amsthm,amssymb,amsfonts}
\usepackage{algorithm}
\usepackage{algorithmic}
\makeatother
\theoremstyle{plain}
\newtheorem{theorem}{Theorem}

\newtheorem{proposition}{Proposition}
\newtheorem{assumption}{Assumption}
\newtheorem{notation}{Notation}
\theoremstyle{definition}

\theoremstyle{remark}

\usepackage{caption} 
\usepackage{fancyhdr}

\author{
  {\normalsize Fr\'ed\'eric Magoul\`es}\thanks{Universit\'e Paris-Saclay, CentraleSup\'elec, Gif-sur-Yvette, France
    (correspondence, frederic.magoules@hotmail.com).}
  \and
  {\normalsize Guillaume Gbikpi-Benissan}\thanks{IRT SystemX, Palaiseau, France
    (guibenissan@gmail.com).}
}
\title{Distributed convergence detection based on global residual error under asynchronous iterations}
\date{}

\begin{document}
\maketitle
\thispagestyle{fancy}

\begin{abstract}
\noindent Convergence of classical parallel iterations is detected by performing a reduction operation at each iteration in order to compute a residual error relative to a potential solution vector. To efficiently run asynchronous iterations, blocking communication requests are avoided, which makes it hard to isolate and handle any global vector. While some termination protocols were proposed for asynchronous iterations, only very few of them are based on global residual computation and guarantee effective convergence. But the most effective and efficient existing solutions feature two reduction operations, which constitutes an important factor of termination delay. In this paper, we present new, non-intrusive, protocols to compute a residual error under asynchronous iterations, requiring only one reduction operation. Various communication models show that some heuristics can even be introduced and formally evaluated. Extensive experiments with up to 5600 processor cores confirm the practical effectiveness and efficiency of our approach.
\end{abstract}

\begin{keywords}
asynchronous iterations; convergence detection; global residual; distributed snapshot; parallel computing
\end{keywords}

\section{Introduction}
\label{sec:intro}

{R}{educing} the impact of communication on the efficiency of a parallel computation is often achieved by optimizing a graph of data dependency between the computing units. However, for iterative methods, another important aspect to take into account is how often data transfers occur. Indeed, in a classical parallel procedure, the computation has to pause each time a remote data is needed due to dependency. This can result in a notable global slowdown of the procedure, according to the properties of the underlying communication platform.
Asynchronous iterations are thus interesting to minimize the impact of this second aspect. By not requiring synchronization at each iteration, asynchronous methods avoid idling while waiting for a data exchange, thus ideally reduce the wasted time. This kind of iterative methods was first experienced in \cite{Rosenfeld:1969:CSP:363626.363628} as part of a study on the simulation of parallel processing. It follows from the asynchronism that the components of a global vector are iteratively computed without the precedence order ensured by synchronous iterations, which obviously introduces some convergence issues.
A first convergence result was established in \cite{Chazan1969199} for the solution of algebraic linear systems, then non linear systems were investigated as well (see, e.g., \cite{Donnelly1971117, Miellou1975}). Performance comparison against synchronous iterations was first conducted on a parallel computer in \cite{Baudet:1978:AIM:322063.322067}.
Many studies confirmed the efficiency of asynchronous methods in various mathematical fields such as the obstacle problem (see, e.g., \cite{bsa1989, SpMiEl2001.3, Chau:2011:PSO:2076556.2076562}), dynamic programming (see, e.g., \cite{Uresin:1990:PAA:79147.79162}), optimization and flow problems (see, e.g., \cite{Chajakis1991873, FLD:FLD1502}), partial differential equations (see, e.g., \cite{doi:10.1080/00207168508803486, Hart1989131}), differential-algebraic systems (see, e.g., \cite{BahiEtAl1996}) Markov chains and optimal control (see, e.g., \cite{jarraya2000}).

Nowadays, asynchronous parallel algorithms are particularly investigated for taking full advantage of massively parallel architectures and largely distributed platforms. Indeed, in such environments, the most part of the efficiency of parallel algorithms relies on the management of interprocess communication. Yet, these new computational environments raise efficiency and accuracy issues about evaluating the convergence state of asynchronous parallel iterative processes. Indeed, with such increasing communication loads, there is no trivial efficient way to compute a consistent residual error from the distributed components of a global, potential solution, vector. Therefore, a well designed detection technique is required, in order to avoid both untimely and delayed termination. In this paper, we propose new efficient methods to accurately evaluate the residual of a computation during asynchronous iterations. Such a matter is not related to conditions under which an asynchronous iterative algorithm is guaranteed to converge, but rather consists in designing some efficient and effective way of asserting that an ongoing asynchronous iterative computation has actually reached its convergence state.

Section~\ref{sec:rw} gives a brief overview of main existing approaches and protocols for terminating asynchronous iterations. Section~\ref{sec:pb} presents the asynchronous iterations model that is under consideration, then formally states the convergence detection problem that is addressed in this study. Section~\ref{sec:snapshot} details basic ideas of snapshot protocols, leading to our propositions for asynchronous iterations termination in First-In-First-Out (FIFO) communication environments. Then, Section~\ref{sec:nfais} tackles various non-FIFO communication contexts. Two new protocols are proposed for arbitrary non-FIFO communication, another one for non-FIFO communication only on messages which have different labels, and at last two others based on heuristics, for non-FIFO communication only within successive finite sets of exchanged messages. Section~\ref{sec:nr} comments some experimental results on two different computation platforms, using up to 5600 processor cores. Effectiveness and efficiency are discussed against two existing termination methods. Section~\ref{sec:conclu} summarizes our conclusions.

\section{Related works}
\label{sec:rw}

The problem of terminating asynchronous iterations was well discussed in, e.g., \cite{Bertsekas19913}, where the authors introduced a first approach which consists in altering the asynchronous iterative algorithm such that it terminates in finite time and then applying one of the classical termination detection protocols available in the distributed algorithms field (see, e.g., \cite{dijkstra1980termination, Francez:1982:ADT:1313330.1313673, Rana198343, Mattern1987}). Indeed, these termination protocols are designed for parallel applications that are executed in a finite number of steps, that is to say, there is a point, during their execution, from where all single processes are idle. Since this is not natively the case for a large class of iterative algorithms, different modifications have been proposed (see, e.g., \cite{Savari199639, doi:10.1080/10637199608915571, chau_algorithmes_2005}) for detecting their convergence by means of a classical distributed termination protocol. Basically, any process under some local conditions (relative to local convergence) stops sending new data to its neighbors in the communication graph, so that the termination condition may consist of having all processes under this local condition, without any message in transit.
Another kind of alteration has been discussed in \cite{doi:10.1080/00207169808804758}, which consists in turning back to synchronous iterations at some point of the execution where local convergence seems to persist on one of the processes.

A second approach, called \textit{supervised termination}, consists in using a supervisory algorithm to take a snapshot of the computation, in order to construct and evaluate a global solution in parallel of the iterative process.
Considering the well-known snapshot protocol due to K. M. Chandy and L. Lamport~\cite{Chandy:1985:DSD:214451.214456}, it is still interesting to see how it applies for asynchronous iterations termination in a simplified form (see Section~\ref{subsec:ais}). Yet, the main disadvantage of such a protocol is the FIFO property required on the communication channels.
Attempts to achieve general non-FIFO snapshots are based either on message acknowledgment and delayed delivering, or on piggybacking of control information on top of application messages (see \cite{0967-1846-2-4-005} for an introductory overview). Such approaches thus turn out to be quite intrusive and, furthermore, not easy to implement.
In \cite{Savari199639}, some supervised termination protocols, more or less centralized, were designed over both star and tree network topologies, introducing a new non-FIFO, but simplified, snapshot. The less centralized approach therein involves a spanning tree over the network graph where local convergence notifications propagate from the leaves to the root process. This one then triggers the simplified snapshot allowing each process to evaluate a globally coherent local solution. The centralization is thus limited to the notifications gathering phase for coordination purpose. Consistency, for non-FIFO channels, is guaranteed by inserting computation message data into the snapshot messages, which introduces a non-negligible overhead for communication.

A third approach in \cite{Bahi:2005:DCD:1038058.1038193} is based on a leader election protocol on tree topology~\cite[Section~4.4.3]{Lynch:1996:DA:525656} wherein the authors introduced cancellation messages to manage the false convergence issue. The algorithm however requires to estimate an upper bound on the communication delay between any two processes.
Then, in \cite{BahiEtAl2008}, these authors proposed a new solution which takes off this requirement, as well as cancellation messages, by performing a verification phase after a presumed global convergence. The leading idea is to monitor the persistence of this convergence state within a period which must last enough to have every dependencies updated with data at least as recent as the presumed detection time. Global convergence is confirmed if during this period no process ever left its local convergence state. As an inconvenient for non-FIFO environments, piggybacking techniques must be used to distinguish data emitted within the verification phase period. While such an approach can avoid premature termination with a high probability, it does not provide a way of evaluating a consistent global residual. Yet, its reliability could be guaranteed by mixing it with the formal analysis from \cite{MielEtAl2008} where the convergence tests are based on the diameter of successive nested sets, which are identified by means of macro-iterations defined as minimal sets of iterations within which all of the solution vector components are updated at least once. Nevertheless, just as in \cite{Savari199639}, this third approach also features a first gathering phase through the leader election, which actually acts as a dynamically centralized coordination.

In summary, second and third approaches allow us to detect the convergence of asynchronous iterations without altering the main computation process. But for both, current solutions somehow require two gathering phases, one for coordination and another for convergence state evaluation. In very large distributed systems, such reduction operations would constitute the most costly part of these convergence detection protocols.
We investigate here new methods, mostly non-intrusive, to evaluate the convergence residual of a computation during asynchronous iterations, using only one reduction operation. Furthermore, some non-FIFO cases are managed through strong heuristics, without piggybacking or over-exchange of computation data.

\section{Problem formulation}
\label{sec:pb}

\subsection{Asynchronous iterations}

Let $X = X_{1} \times \cdots \times X_{n}$ be a product of vector spaces, and let us consider a mapping
\[
\begin{array}{lccc}
f : & X_{1} \times \cdots \times X_{n} & \to & X_{1} \times \cdots \times X_{n},\\
& x = (x_{1}, \ldots, x_{n}) & \mapsto & (f_{1}(x), \ldots, f_{n}(x)),
\end{array}
\]
where $f_{i} : X \to X_{i}$, $i \in \{1, \ldots, n\}$, are given. Now let $\{I^{k}\}_{k \in \mathbb{N}}$ be a sequence of integer subsets such that
\[
\forall k \in \mathbb{N}, \quad I^{k} \subseteq \{1, \ldots, n\}, \quad I^{k} \ne \emptyset.
\]
Asynchronous iterations exhibit a sequence $\{x^{k}\}_{k \in \mathbb{N}}$ of vectors in $X$ such that
\begin{equation}
\label{eq:ai}
x_{i}^{k+1} = \left \{
\begin{array}{ll}
f_{i}(x_{1}^{\rho_{1}^{i}(k)}, \ldots, x_{n}^{\rho_{n}^{i}(k)}), & i \in I^{k},\\
x_{i}^{k}, & i \notin I^{k},
\end{array}
\right.
\end{equation}
where $\rho_{j}^{i}$, with $i, j \in \{1, \ldots, n\}$, are integer-valued functions on $\mathbb{N}$, satisfying
\[
\rho_{j}^{i}(k) \le k, \quad \forall k \in \mathbb{N},
\]
which denotes a delay on the version of the component $j$ used to update the component $i$. $I^{k}$ is thus the set of components updated at iteration $k$.
For convergence analysis, the computational model \eqref{eq:ai} is generally completed by the two following assumptions, which ensure that, for any given $k_{0} \in \mathbb{N}$ before convergence, no component sequence $\{x_{i}^{k_{0}}, x_{i}^{k_{0}+1}, \ldots\}$ definitively freezes or is generated by using some other fixed component $x_{j}^{k_{1}}$, $k_{1} \in \mathbb{N}$.
\begin{assumption}
\label{ass:ai_comp}
$\forall i \in \{1, \dots, n\}$, $\operatorname{card}\{k \in \mathbb{N} | i \in I^{k}\} = +\infty$.
\end{assumption}
\begin{assumption}
\label{ass:ai_comm}
$\forall i,j \in \{1, \dots, n\}$, $\underset{k \to +\infty}{\lim} \rho_{j}^{i}(k) = +\infty$.
\end{assumption}

\subsection{Convergence detection}
\label{subsec:cvg_detect}

Let us consider a sequence $\{x^{k}\}_{k \in \mathbb{N}}$ of vectors in $X$ satisfying the asynchronous iterations model \eqref{eq:ai}, and define $n$ sequences $\{y^{1,k}\}_{k \in \mathbb{N}}, \ldots, \{y^{n,k}\}_{k \in \mathbb{N}}$ of vectors in $X$ such that
\begin{equation}
\label{eq:ai2}
y^{i,k} = (x_{1}^{\rho_{1}^{i}(k)}, \ldots, x_{n}^{\rho_{n}^{i}(k)}), \quad \forall i \in \{1, \ldots, n\}, \forall k \in \mathbb{N}.
\end{equation}
$y^{i,k}$ thus denotes the global vector used to update the component $i$ of the solution vector $x$ at the iteration $k+1$.
Additionally, we assume to have
\begin{equation}
\label{eq:ai3}
\rho_{i}^{i}(k) = k, \quad \forall i \in \{1, \ldots, n\}, \forall k \in \mathbb{N}.
\end{equation}
At last, let $\bar{x}$ be a vector in $X$ given by
\[
\bar{x} = (y_{1}^{1,k_{1}}, \ldots, y_{n}^{n,k_{n}}), \quad k_{1}, \ldots, k_{n} \in \mathbb{N},
\]
which denotes a global vector built from an arbitrary version of each local component.
We will address in this paper the problem of evaluating a relation
\begin{equation}
\label{eq:prob}
\| f(\bar{x}) - \bar{x} \| < \varepsilon, \quad \varepsilon \in \mathbb{R},
\end{equation}
where $\|.\|$ is a norm on $X$. To solve this problem, attention will be mainly paid to the computation of $f(\bar{x})$.

One notices that synchronous iterations correspond to the case where we have $I^{k} = \{1, \ldots, n\}$ and $\rho_{j}^{i}(k) = k$, for all $k \in \mathbb{N}$ and all $i, j \in \{1, \ldots, n\}$. It follows that by taking $\bar{x} = (y_{1}^{1,k}, \ldots, y_{n}^{n,k})$, for any $k \in \mathbb{N}$, we obtain, for all $i \in \{1, \ldots, n\}$,
\[
\begin{aligned}
f_{i}(\bar{x}) & = f_{i}(y_{1}^{1,k}, \ldots, y_{n}^{n,k}),\\
& = f_{i}(x_{1}^{k}, \ldots, x_{n}^{k}),\\
& = f_{i}(x_{1}^{\rho_{1}^{i}(k)}, \ldots, x_{n}^{\rho_{n}^{i}(k)}),\\
& = x_{i}^{k+1},
\end{aligned}
\]
which implicitly gives $f(\bar{x}) = (y_{1}^{1,k+1}, \ldots, y_{n}^{n,k+1})$.

We point out that the relation~\eqref{eq:prob} can be more generally given by:
\begin{equation}
\label{eq:prob_gen}
\bar{x} \in S^{*},
\end{equation}
where $S^{*}$ is the set of admissible solutions, as also suggested in \cite{Savari199639}. Then actually, the quality of the solution $\bar{x}$ would depend on the suitable choice of a residual evaluation function $r(\bar{x})$, which is application-dependent, regardless the context of asynchronous iterations. By considering however a residual evaluation function of the form~\eqref{eq:prob}, we intend to provide a better understanding of the subsequent discussions, without losing their general applicability to~\eqref{eq:prob_gen}.

\section{Determining a global solution vector}
\label{sec:snapshot}

\subsection{The Chandy--Lamport snapshot (CLS)}

The basic idea within the CLS protocol is to record, not only the local state of each process, but also the state of each communication channel. Any process (possibly several processes) can initiate the protocol by recording its local state and sending a ``marker" to all of its neighbors in the communication graph. Non-initiators do the same when they receive a marker for the first time. As soon as a process records its local state, it starts recording the state of its reception channels. From then, and before marker reception on any channel, any message received is appended to the state of this channel. Consequently, the recording ends when a marker is received from all of the neighboring processes. Algorithm~\ref{algo:cls} outlines the rules that fully describe the protocol.
\begin{algorithm}[!t]
\caption{CLS protocol}
\label{algo:cls}
\begin{algorithmic}[1]
\IF {initiator}
	\IF {state not recorded}
		\STATE{Record state}
		\STATE{Send a marker to each neighbor in the communication graph}
	\ENDIF
\ENDIF
\IF {marker received}
	\IF {state not recorded}
		\STATE{Record state}
		\STATE{Send a marker to each neighbor in the communication graph}
	\ENDIF
	\IF {marker received from each neighbor}
		\STATE{Return state and state of each reception channel}
	\ENDIF
\ENDIF
\IF {computation message received}
	\IF {state recorded and marker not received from the sender}
		\STATE{Add the message to the state of the corresponding reception channel}
	\ENDIF
\ENDIF
\end{algorithmic}
\end{algorithm}

\begin{figure}[!t]
\centering
\includegraphics[scale=0.35]{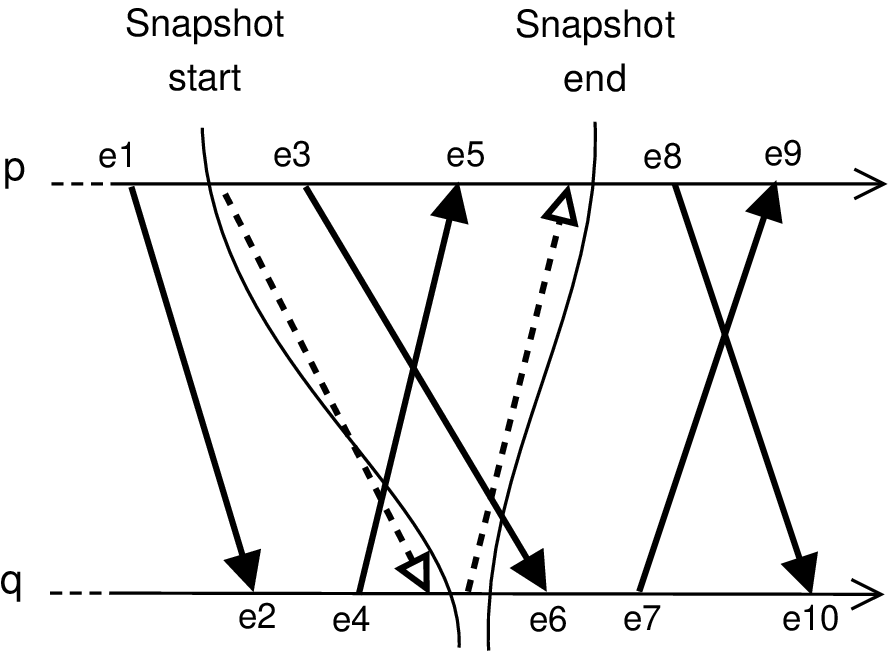}
\caption{Example of a CLS protocol execution with two processes.}
\label{fig:cls}
\end{figure}

To give an intuitive understanding of the consistency of the global state built by this snapshot protocol, we show, in Figure~\ref{fig:cls}, a simple example involving two processes, denoted by $p$ and $q$.
Let us consider events consisting in sending and receiving a message. In this example, the process $p$ records its local state after the event $e1$ and sends a marker (dotted arrow) to the process $q$. On reception of the marker, the process $q$ records its local state after the event $e4$, then records the state of its reception channel as an empty set, and finally, sends the marker back to the process $p$. Before receiving the marker from the process $q$, the process $p$ received a computation message from $q$ as event $e5$. Therefore, the state of the reception channel of the process $p$ corresponds to the set $\{e5\}$. It is clear from this example that the communication channels need to be FIFO. Otherwise, if for instance the marker sent by the process $q$ is received by the process $p$ before the event $e5$, therefore the state of the channel is an empty set, which causes an information lost about the event $e5$.

This example builds a global state relative to last events $\{e1, e4\}$ and records the set of pending messages relative to event $e5$. However, according to the events sequence, this state does not match any of the states the system actually went through. Indeed, one can see that the event $e3$ should be taken into account as we consider the state of the system just after the event $e4$. Therefore, let us highlight what is relevant about the state recorded by an execution of the CLS protocol.
\begin{theorem}[Chandy \& Lamport, 1985]
\label{theo:cls}
Let $\mathcal{S(C)} = \{s^{t}\}_{t \in \mathbb{N}}$ denote the global states sequence generated by a computation $\mathcal{C}$. Let $\bar{s}$ be the global state recorded by an execution of the CLS protocol on $\mathcal{C}$. Then there exists an equivalent permutation $\mathcal{P(C)}$ of $\mathcal{C}$ such that $\bar{s} \in \mathcal{S(P(C))}$.
\end{theorem}
\begin{proof}
See~\cite{Chandy:1985:DSD:214451.214456}.
\end{proof}

\subsection{New asynchronous iterations snapshots (AIS)}
\label{subsec:ais}

Let again sequences $\{x^{k}\}_{k \in \mathbb{N}}$ and $\{y^{i,k}\}_{k \in \mathbb{N}}$, $i \in \{1, \ldots, n\}$, be defined as in Section~\ref{subsec:cvg_detect}. Let us suppose an associated parallel computation involving $n$ processes, and let each process $i \in \{1, \ldots, n\}$ record a vector $\bar{y}^{i} \in X$ by following the rules described either in Algorithm~\ref{algo:ais1} or in Algorithm~\ref{algo:ais2}.
\begin{algorithm}[!t]
\caption{AIS protocol 1}
\label{algo:ais1}
\begin{algorithmic}[1]
\IF {$\|y_{i}^{i,k} - y_{i}^{i,k_{0}^{i}}\| < \varepsilon$, with $y_{i}^{i,k} = y_{i}^{i,k_{0}^{i}+1}$, $i \in I^{k_{0}^{i}}$}
	\IF {$\bar{y}_{i}^{i}$ undefined}
		\STATE{$\bar{y}_{i}^{i} := y_{i}^{i,k}$}
		\FORALL {process $j \ne i$}
			\STATE{Send a marker to $j$}
		\ENDFOR
	\ENDIF
\ENDIF
\IF {marker received from a process $j \ne i$}
	\STATE{$\bar{y}_{j}^{i} := y_{j}^{i,k}$}
	\IF {$\bar{y}_{i}^{i}$ undefined}
		\STATE{$\bar{y}_{i}^{i} := y_{i}^{i,k}$}
		\FORALL {process $j \ne i$}
			\STATE{Send a marker to $j$}
		\ENDFOR
	\ENDIF
	\IF {$\bar{y}_{j}^{i}$ defined for all $j$}
		\RETURN{$\bar{y}^{i}$}
	\ENDIF
\ENDIF
\end{algorithmic}
\end{algorithm}
\begin{algorithm}[!t]
\caption{AIS protocol 2}
\label{algo:ais2}
\begin{algorithmic}[1]
\IF {$\|y_{i}^{i,k} - y_{i}^{i,k_{0}^{i}}\| < \varepsilon$, with $y_{i}^{i,k} = y_{i}^{i,k_{0}^{i}+1}$, $i \in I^{k_{0}^{i}}$}
	\IF {$\bar{y}_{i}^{i}$ undefined}
		\STATE{$\bar{y}_{i}^{i} := y_{i}^{i,k}$}
		\FORALL {process $j \ne i$}
			\STATE{Send a marker to $j$}
		\ENDFOR
	\ENDIF
\ENDIF
\IF {marker received from a process $j \ne i$}
	\STATE{$\bar{y}_{j}^{i} := y_{j}^{i,k}$}
\ENDIF
\IF {$\bar{y}_{j}^{i}$ defined for all $j$}
	\RETURN{$\bar{y}^{i}$}
\ENDIF
\end{algorithmic}
\end{algorithm}
We should mention that the variable $k$ therein may have different values at different places in the algorithms, as the rules conditions may be fulfilled at different times. To be more precise, we would then have
\[
\bar{y}^{i} = (y_{1}^{i,k_{i,1}}, \ldots, y_{n}^{i,k_{i,n}}), \quad\quad k_{i,j} \in \mathbb{N}, \quad i,j \in \{1, \ldots, n\}.
\]
One can notice that, contrarily to the CLS protocol, there is no rule for channel record at computation message reception. More, in Algorithm~\ref{algo:ais2}, recording the local state is not required at the first marker reception. However, for both algorithms, we still need the following preliminary assumptions.
\begin{assumption}
\label{ass:ais1}
Each process performs at least one iteration, which means :
\[
\forall i \in \{1, \ldots, n\}, \quad \exists k < k_{i,i} : i \in I^{k}.
\]
\end{assumption}
\begin{assumption}
\label{ass:ais2}
After computation of $y_{i}^{i,k+1}$ (i.e., $i \in I^{k}$), $y_{i}^{i,k+1}$ is sent to each process $j \ne i$, before any other communication toward $j$.
\end{assumption}
\begin{assumption}
\label{ass:ais3}
Communication channels are FIFO.
\end{assumption}
A consistent global solution vector, under asynchronous iterations, is then given by the following result.
\begin{proposition}
\label{prop:ais}
Let a sequence $\{x^{k}\}_{k \in \mathbb{N}}$, satisfying the asynchronous iterations model \eqref{eq:ai}, be generated by a computation $\mathcal{C}$ involving $n$ processes. Let sequences $\{y^{1,k}\}_{k \in \mathbb{N}}, \ldots, \{y^{n,k}\}_{k \in \mathbb{N}}$ be defined by the rewriting \eqref{eq:ai2}. Let, at last, $\bar{y}^{1}, \ldots, \bar{y}^{n}$ be the vectors returned by an execution of either the AIS protocol 1 or the AIS protocol 2 on $\mathcal{C}$. Then, under Assumptions~\ref{ass:ais1} to~\ref{ass:ais3}, we have
\[
\bar{y}^{1} = \bar{y}^{2} = \cdots = \bar{y}^{n}.
\]
\end{proposition}
\begin{proof}
Let $i, j \in \{1, \ldots, n\}$ be two any process identifiers. According to the local state recording rule and Assumption~\ref{ass:ais1}, there exists $k_{0}^{i} < k_{i,i}$, with $i \in I^{k_{0}^{i}}$, satisfying :
\[
\forall k \in \{k_{0}^{i}+1, \ldots, k_{i,i}-1\}, \quad i \notin I^{k},
\]
so that we have
\[
y_{i}^{i,k_{i,i}} = y_{i}^{i,k_{i,i}-1} = \cdots = y_{i}^{i,k_{0}^{i}+1}.
\]
With Assumptions~\ref{ass:ais2} and~\ref{ass:ais3}, it follows that there also exists $k_{0}^{j} \in \mathbb{N}$, $k_{0}^{j} \le k_{j,i}$, such that
\[
y_{i}^{j,k_{0}^{j}} = y_{i}^{i,k_{0}^{i}+1}.
\]
Assumption~\ref{ass:ais3} implies that
\[
\forall k \in \{k_{0}^{j}+1, \ldots, k_{j,i}\}, \quad y_{i}^{j,k} = y_{i}^{j,k_{0}^{j}}.
\]
Then, in particular, we have
\[
y_{i}^{j,k_{j,i}} = y_{i}^{j,k_{0}^{j}} = y_{i}^{i,k_{0}^{i}+1} = y_{i}^{i,k_{i,i}},
\]
and thus
\begin{equation}
\label{eq:fifo}
\bar{y}_{i}^{i} = \bar{y}_{i}^{j}, \quad \forall i,j \in \{1, \ldots, n\},
\end{equation}
which concludes the proof.
\end{proof}
We can thus have a vector $\bar{x} = (\bar{y}_{1}^{1}, \ldots, \bar{y}_{n}^{n})$, so that we implicitly obtain
\[
f(\bar{x}) = (f_{1}(\bar{y}^{1}), \ldots, f_{n}(\bar{y}^{n})).
\]
Assumptions~\ref{ass:ais1} and~\ref{ass:ais2} are pretty natural conditions that are easily satisfied in an iterative loop where the AIS protocol rules are called after the main computation and message sending part. They are necessary to be mentioned, however, especially for multi-threaded processes. Assumption~\ref{ass:ais3} is then the sole actual constraint in the above protocols. The next section discusses about taking off such requirement.

\section{New non-FIFO asynchronous iterations snapshots}
\label{sec:nfais}

\subsection{Arbitrary non-FIFO communication}

\begin{figure}[!t]
\centering
\includegraphics[scale=0.35]{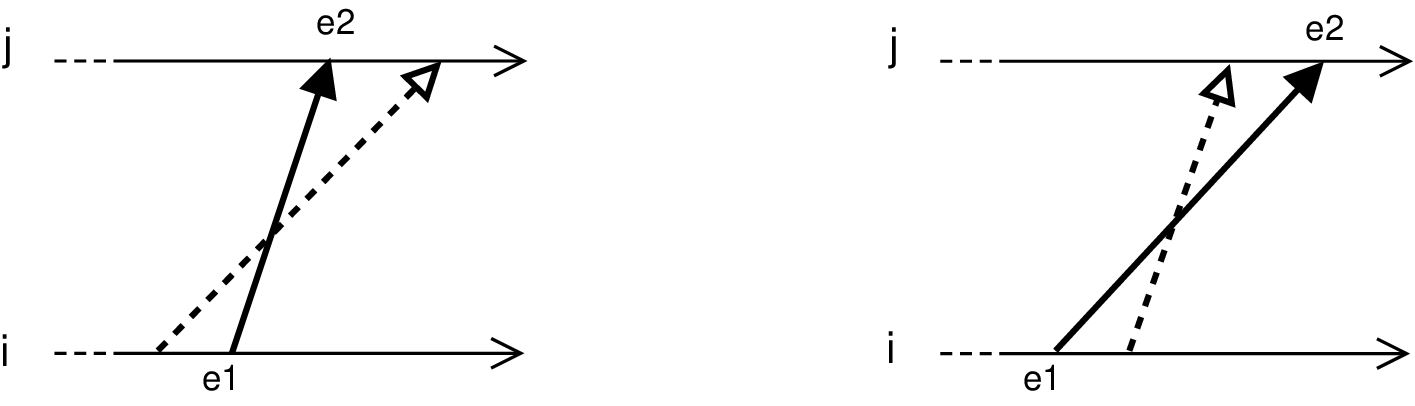}
\caption{Non-FIFO snapshot issues.}
\label{fig:fifo}
\end{figure}

The FIFO condition is essential to avoid the two situations depicted in Figure~\ref{fig:fifo}, where a marker (dotted arrow) crosses a computation message.
In such cases, the equality in~\eqref{eq:fifo} is no more satisfied. Then, one can apply ideas from~\cite{Savari199639} to AIS protocols 1 and 2, as described by Algorithms~\ref{algo:nf_ais1} and~\ref{algo:nf_ais2}, respectively.
\begin{algorithm}[!t]
\caption{Non-FIFO AIS protocol~1}
\label{algo:nf_ais1}
\begin{algorithmic}[1]
\IF {$\|y_{i}^{i,k} - y_{i}^{i,k_{0}^{i}}\| < \varepsilon$, with $y_{i}^{i,k} = y_{i}^{i,k_{0}^{i}+1}$, $i \in I^{k_{0}^{i}}$}
	\IF {$\bar{y}_{i}^{i}$ undefined}
		\STATE{$\bar{y}_{i}^{i} := y_{i}^{i,k}$}
		\FORALL {process $j \ne i$}
			\STATE{Send a marker $\bar{y}_{i}^{i}$ to $j$}
		\ENDFOR
	\ENDIF
\ENDIF
\IF {marker $\bar{y}_{j}^{j}$ received from a process $j \ne i$}
	\STATE{$\bar{y}_{j}^{i} := \bar{y}_{j}^{j}$}
	\IF {$\bar{y}_{i}^{i}$ undefined}
		\STATE{$\bar{y}_{i}^{i} := y_{i}^{i,k}$}
		\FORALL {process $j \ne i$}
			\STATE{Send a marker $\bar{y}_{i}^{i}$ to $j$}
		\ENDFOR
	\ENDIF
	\IF {$\bar{y}_{j}^{i}$ defined for all $j$}
		\RETURN{$\bar{y}^{i}$}
	\ENDIF
\ENDIF
\end{algorithmic}
\end{algorithm}
\begin{algorithm}[!t]
\caption{Non-FIFO AIS protocol~2}
\label{algo:nf_ais2}
\begin{algorithmic}[1]
\IF {$\|y_{i}^{i,k} - y_{i}^{i,k_{0}^{i}}\| < \varepsilon$, with $y_{i}^{i,k} = y_{i}^{i,k_{0}^{i}+1}$, $i \in I^{k_{0}^{i}}$}
	\IF {$\bar{y}_{i}^{i}$ undefined}
		\STATE{$\bar{y}_{i}^{i} := y_{i}^{i,k}$}
		\FORALL {process $j \ne i$}
			\STATE{Send a marker $\bar{y}_{i}^{i}$ to $j$}
		\ENDFOR
	\ENDIF
\ENDIF
\IF {marker $\bar{y}_{j}^{j}$ received from a process $j \ne i$}
	\STATE{$\bar{y}_{j}^{i} := \bar{y}_{j}^{j}$}
\ENDIF
\IF {$\bar{y}_{j}^{i}$ defined for all $j$}
	\RETURN{$\bar{y}^{i}$}
\ENDIF
\end{algorithmic}
\end{algorithm}
Here, markers contain computation data, so that these solutions actually even handle crossed computation messages. Proposition~\ref{prop:ais} becomes the following, which does not need any of the previous assumptions :
\begin{proposition}
\label{prop:nf_ais12}
Let a sequence $\{x^{k}\}_{k \in \mathbb{N}}$, satisfying the asynchronous iterations model \eqref{eq:ai}, be generated by a computation $\mathcal{C}$ involving $n$ processes. Let sequences $\{y^{1,k}\}_{k \in \mathbb{N}}, \ldots, \{y^{n,k}\}_{k \in \mathbb{N}}$ be defined by the rewriting \eqref{eq:ai2}. Let, at last, $\bar{y}^{1}, \ldots, \bar{y}^{n}$ be the vectors returned by an execution of either the non-FIFO AIS protocol 1 or the non-FIFO AIS protocol 2 on $\mathcal{C}$. Then, we have
\[
\bar{y}^{1} = \bar{y}^{2} = \cdots = \bar{y}^{n}.
\]
\end{proposition}
\begin{proof}
By construction, we trivially satisfy the equality in~\eqref{eq:fifo}.
\end{proof}

\subsection{Inter-protocol non-FIFO communication}

In the communication model considered now, FIFO channels are used at least for computation messages. This is a highly realistic model, as being a natural expectation to achieve minimum delays during asynchronous iterations, and moreover, it is a requirement for classical iterations. Still, the problem of markers crossing computation messages remains. We propose, with Algorithm~\ref{algo:nf_ais3}, a snapshot solution which, outrightly, do not need marker exchange, and is based on only computation messages, even without piggybacking.
\begin{algorithm}[!t]
\caption{Non-FIFO AIS protocol~3}
\label{algo:nf_ais3}
\begin{algorithmic}[1]
\IF {$\|y_{i}^{i,k} - y_{i}^{i,k_{i}^{i}}\| < \varepsilon$, with $y_{i}^{i,k} = y_{i}^{i,k_{i}^{i}+1}$, $i \in I^{k_{i}^{i}}$}
	\IF {$\bar{y}_{i}^{i}$ undefined}
		\STATE{$\bar{y}_{i}^{i} := y_{i}^{i,k}$}
	\ENDIF
\ENDIF
\IF {$\|y_{j}^{i,k} - y_{j}^{i,k_{j}^{i}}\| < \varepsilon$, with $\rho_{j}^{i}(k_{j}^{i}) = k_{j}^{j}$, $\rho_{j}^{i}(k) = k_{j}^{j}+1$}
	\IF {$\bar{y}_{j}^{i}$ undefined}
		\STATE{$\bar{y}_{j}^{i} := y_{j}^{i,k}$}
	\ENDIF
\ENDIF
\IF {$\bar{y}_{j}^{i}$ defined for all $j$}
	\RETURN{$\bar{y}^{i}$}
\ENDIF
\end{algorithmic}
\end{algorithm}
Here, just as local solution buffers, each process $i$ maintains access to the two latest received messages, for all neighbor processes $j \ne i$. Then, process $i$ can detect by itself local convergence of process $j$ and immediately record the last value received. Proposition~\ref{prop:ais} becomes the following :
\begin{proposition}
\label{prop:nf_ais3}
Let a sequence $\{x^{k}\}_{k \in \mathbb{N}}$, satisfying the asynchronous iterations model \eqref{eq:ai}, be generated by a computation $\mathcal{C}$ involving $n$ processes. Let sequences $\{y^{1,k}\}_{k \in \mathbb{N}}, \ldots, \{y^{n,k}\}_{k \in \mathbb{N}}$ be defined by the rewriting \eqref{eq:ai2}. Let, at last, $\bar{y}^{1}, \ldots, \bar{y}^{n}$ be the vectors returned by an execution of the non-FIFO AIS protocol 3 on $\mathcal{C}$. Then, we have
\[
\bar{y}^{1} = \bar{y}^{2} = \cdots = \bar{y}^{n}.
\]
\end{proposition}
\begin{proof}
Let $i, j \in \{1, \ldots, n\}$ be two any process identifiers. Remind $\bar{y}_{j}^{i} = y_{j}^{i, k_{i,j}}$, $k_{i,j} \in \mathbb{N}$. Then according to~\eqref{eq:ai2} and~\eqref{eq:ai3}, we have
\[
\bar{y}_{j}^{i} = x_{j}^{\rho_{j}^{i}(k_{i,j})} = x_{j}^{\rho_{j}^{j}(\rho_{j}^{i}(k_{i,j}))} = y_{j}^{j, \rho_{j}^{i}(k_{i,j})}.
\]
By construction, we satisfy
\[
\rho_{j}^{i}(k_{i,j}) = k_{j}^{j} + 1, \quad\quad y_{j}^{j, k_{j,j}} = y_{j}^{j, k_{j}^{j}+1},
\]
and thus
\[
\bar{y}_{j}^{i} = y_{j}^{j, k_{j}^{j} + 1} = y_{j}^{j, k_{j,j}} = \bar{y}_{j}^{j},
\]
which concludes the proof.
\end{proof}

\subsection{Non-FIFO communication with bounded number of cross messages}

\begin{figure}[!t]
\centering
\includegraphics[scale=0.33]{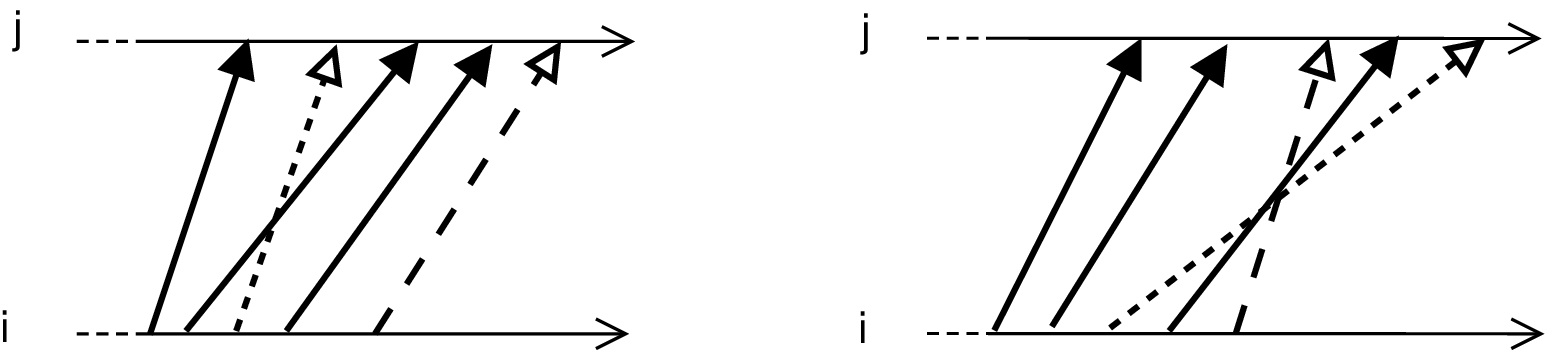}
\caption{Examples of issues handled by non-FIFO AIS protocol 4.}
\label{fig:ais}
\end{figure}

In case of very large problems, non-FIFO AIS protocols 1 to 3 may introduce non-negligible overhead costs, either for communication or for memory. But on another hand, for such large problems, deciding to compute a solution may depend on guaranteeing a minimum performance level of the parallel computation platform. Especially, when a given maximum execution time is expected, this most likely includes to ensure a bound on communication delays. We thus reasonably make here a preliminary assumption.
\begin{assumption}
\label{ass:sf}
A message can cross at most $\eta$ other messages.
\end{assumption}
Let us then consider Algorithm~\ref{algo:nf_ais4}.
\begin{algorithm}[!t]
\caption{Non-FIFO AIS protocol~4}
\label{algo:nf_ais4}
\begin{algorithmic}[1]
\IF {$\|y_{i}^{i,t+1} - y_{i}^{i,t}\| < \varepsilon, \forall t \in \{k_{0}^{i}, \ldots, k-1\} : i \in I^{t}$}
	\IF {$\bar{y}_{i}^{i}$ undefined}
		\STATE{$\bar{y}_{i}^{i} := y_{i}^{i,k}$}
		\FORALL {process $j \ne i$}
			\STATE{Send a marker to $j$}
		\ENDFOR
		\STATE{$k_{i,i} := k$}
		\STATE{Mark $\phi_{i}^{i}$ as undefined}
	\ENDIF
\ENDIF
\IF {$\|y_{i}^{i,t+1} - y_{i}^{i,t}\| < \varepsilon, \forall t \in \{k_{i,i}, \ldots, k-1\} : i \in I^{t}$}
	\IF {$\phi_{i}^{i}$ undefined}
		\STATE{$\phi_{i}^{i} := 1$}
		\FORALL {process $j \ne i$}
			\STATE{Send a flagged marker $\phi_{i}^{i}$ to $j$}
		\ENDFOR
	\ENDIF
\ELSE
	\IF {$\phi_{i}^{i}$ undefined}
		\STATE{$\phi_{i}^{i} := 0$}
		\FORALL {process $j \ne i$}
			\STATE{Send a flagged marker $\phi_{i}^{i}$ to $j$}
		\ENDFOR
		\STATE{Mark $\bar{y}_{i}^{i}$ as undefined}
	\ENDIF
\ENDIF
\IF {marker received from a process $j \ne i$}
	\STATE{$\bar{y}_{j}^{i} := y_{j}^{i,k}$}
\ENDIF
\IF {flagged marker $\phi_{j}^{j}$ received from a process $j \ne i$}
	\STATE{$\phi_{j}^{i} := \phi_{j}^{j}$}
	\IF {$\phi_{j}^{i} = 0$}
		\STATE{Mark $\bar{y}_{j}^{i}$ as undefined}
	\ENDIF
\ENDIF
\IF {$\bar{y}_{j}^{i}$ defined and $\phi_{j}^{i} = 1$ for all $j$}
	\RETURN{$\bar{y}^{i}$}
\ENDIF
\end{algorithmic}
\end{algorithm}
Here, a process $i$ sends its marker to a process $j \ne i$ only when local convergence persists on process $i$ for some iterations $k_{l}^{i}$, with $i \in I^{k_{l}^{i}}$ and $l \in \mathbb{N}$. Such iterations will be referred to as `steady iterations'. This way, even if the marker is received on the process $j$ before the latest message sent by the process $i$, the message recorded by the process $j$ is still relevant in the sense that the two latest messages from process $i$ contain very close data (due to the persistence of the local convergence). Then, a second type of marker (dashed arrow in Figure~\ref{fig:ais}) is sent by the process $i$ to transmit a binary flag after some additional iterations. If local convergence still persists during these iterations, the flag is armed, which confirms the relevance of the message data recorded by the process $j$, even if it corresponds to the message sent by the process $i$ after the first marker (again, due to local convergence persistence after sending the first marker). Otherwise, processes $i$ and $j$ discard the corresponding records and try again. One can also see that the algorithm still works even in the case where the flag-marker crosses the first one, as depicted in Figure~\ref{fig:ais} (right).

As a particular case of this communication model, one may further assume that the crossing ability is tightly related to the size of the messages. Indeed, if control messages (e.g., markers) are transmitted far faster than computation messages (due to the difference in size), we may assume that, from a process to another process, a computation message sent later than a control message cannot be received earlier than this one. Then in such case, flagged markers would not be necessary any more, which rather simplifies the protocol and provides Algorithm~\ref{algo:nf_ais5}.
\begin{algorithm}[!t]
\caption{Non-FIFO AIS protocol~5}
\label{algo:nf_ais5}
\begin{algorithmic}[1]
\IF {$\|y_{i}^{i,t+1} - y_{i}^{i,t}\| < \varepsilon, \forall t \in \{k_{0}^{i}, \ldots, k-1\} : i \in I^{t}$}
	\IF {$\bar{y}_{i}^{i}$ undefined}
		\STATE{$\bar{y}_{i}^{i} := y_{i}^{i,k}$}
		\FORALL {process $j \ne i$}
			\STATE{Send a marker to $j$}
		\ENDFOR
	\ENDIF
\ENDIF
\IF {marker received from a process $j \ne i$}
	\STATE{$\bar{y}_{j}^{i} := y_{j}^{i,k}$}
\ENDIF
\IF {$\bar{y}_{j}^{i}$ defined for all $j$}
	\RETURN{$\bar{y}^{i}$}
\ENDIF
\end{algorithmic}
\end{algorithm}

Now, let us define the mapping
\[
\begin{array}{lccc}
g : & X^{n} & \to & X_{1} \times \cdots \times X_{n},\\
& (y^{1}, \ldots, y^{n}) & \mapsto & (f_{1}(y^{1}), \ldots, f_{n}(y^{n})),
\end{array}
\]
and the vector $\bar{y} = (\bar{y}^{1}, \ldots, \bar{y}^{n})$, so that we implicitly obtain
\[
g(\bar{y}) = (f_{1}(\bar{y}^{1}), \ldots, f_{n}(\bar{y}^{n})).
\]
In the following, we establish the reliability of the approximated residual
\[
\|g(\bar{y}) - \bar{x}\|, \quad \bar{x} = (\bar{y}_{1}^{1}, \ldots, \bar{y}_{n}^{n}),
\]
compared to the exact one given by $\|f(\bar{x}) - \bar{x}\|$. Let then $\|.\|_{(i)}$, $i \in \{1, \ldots, n\}$, be a given norm defined on $X_{i}$, and let us consider $\mathcal{L}_{p}$-norms, $p \in [1, +\infty)$, defined on $X$ by
\[
\|x\|_{p} = \left( \sum_{i = 1}^{n} {\|x_{i}\|_{(i)}}^{p} \right)^{1/p}.
\]
Maximum norms could be considered as well, as particular cases. We assume the following property for the mapping $f$.
\begin{assumption}
\label{ass:var_coef}
For any $i$ and $j$ in $\{1, \ldots, n\}$, there exists $\delta_{i,j}$ in $\mathbb{R}^{+*}$ such that :
\[
\|x_{j} - {x'}_{j}\|_{(j)} < \varepsilon
\]
implies
\[
\|f_{i}(x) - f_{i}(x_{1}, \ldots, {x'}_{j}, \ldots, x_{n})\|_{(i)} < \delta_{i,j} \varepsilon,
\]
with $x$ and $x'$ in $X$.
\end{assumption}
\begin{notation}
\label{not:max_var_coef}
$
\displaystyle \delta(f) = \max_{i=1}^{n} \sum_{j=1}^{n} \delta_{i,j}(f),
$
where $\delta_{i,j}(f)$ are the smallest $\delta_{i,j}$ satisfying Assumption~\ref{ass:var_coef}.
\end{notation}
At last, we also need the following assumption.
\begin{assumption}
\label{ass:nf_ais}
A process sends its markers and armed flag-markers after at least $\eta$ steady iterations.
\end{assumption}
Then, we give an essential result about the accuracy of our heuristics.
\begin{proposition}
\label{prop:nf_ais4}
Let a sequence $\{x^{k}\}_{k \in \mathbb{N}}$, satisfying the asynchronous iterations model \eqref{eq:ai}, be generated by a computation $\mathcal{C}$ involving $n$ processes. Let sequences $\{y^{1,k}\}_{k \in \mathbb{N}}, \ldots, \{y^{n,k}\}_{k \in \mathbb{N}}$ be defined by the rewriting \eqref{eq:ai2}. Let, at last, $\bar{y}^{1}, \ldots, \bar{y}^{n}$ be the vectors returned by an execution of the non-FIFO AIS protocol 4 on $\mathcal{C}$. Then, under Assumptions~\ref{ass:ais2} and~\ref{ass:sf} to~\ref{ass:nf_ais}, we have
\[
\|f(\bar{x}) - \bar{x}\|_{p} - \|g(\bar{y}) - \bar{x}\|_{p} < n^{1/p} \eta \delta(f) \varepsilon,
\]
with $\bar{y} = (\bar{y}^{1}, \ldots, \bar{y}^{n})$ and $\bar{x} = (\bar{y}_{1}^{1}, \ldots, \bar{y}_{n}^{n})$.
\end{proposition}
\begin{proof}
Let us take again
\[
\bar{y}_{j}^{i} = y_{j}^{i, k_{i,j}}, \quad \forall i, j \in \{1, \ldots, n\},
\]
with $k_{i,j} \in \mathbb{N}$. Then according to \eqref{eq:ai2} and \eqref{eq:ai3}, we have
\[
\bar{y}_{i}^{j} = y_{i}^{j, k_{j,i}} = x_{i}^{\rho_{i}^{j}(k_{j,i})} = x_{i}^{\rho_{i}^{i}(\rho_{i}^{j}(k_{j,i}))} = y_{i}^{i, \rho_{i}^{j}(k_{j,i})}.
\]
Assumptions~\ref{ass:ais2},~\ref{ass:sf} and~\ref{ass:nf_ais} ensure
\begin{equation}
\label{eq:max_m}
\left| \left\{ k \in \{ \rho_{i}^{j}(k_{j,i}), \ldots, k_{i,i}-1 \}~|~i \in I^{k} \right\} \right| \le \eta.
\end{equation}
Let us then consider
\[
\{k_{1}^{i}, \ldots, k_{m_{i}}^{i}\} = \left\{ k \in \{ \rho_{i}^{j}(k_{j,i}), \ldots, k_{i,i}-1 \}~|~i \in I^{k} \right\},
\]
with $m_{i} \in \mathbb{N}^{*}$. It follows
\[
\begin{aligned}
\|\bar{y}_{i}^{i} - \bar{y}_{i}^{j}\|_{(i)} & = && \|y_{i}^{i, k_{i,i}} - y_{i}^{i, \rho_{i}^{j}(k_{j,i})}\|_{(i)},\\
& = && \|y_{i}^{i, k_{m_{i}}^{i}+1} - y_{i}^{i, k_{1}^{i}}\|_{(i)},\\
& = && \|y_{i}^{i, k_{m_{i}}^{i}+1} - y_{i}^{i, k_{m_{i}}^{i}} + y_{i}^{i, k_{m_{i}}^{i}} - y_{i}^{i, k_{m_{i}-1}^{i}}\\
&&& + \cdots + y_{i}^{i, k_{2}^{i}} - y_{i}^{i, k_{1}^{i}}\|_{(i)},\\
& \le && \|y_{i}^{i, k_{m_{i}}^{i}+1} - y_{i}^{i, k_{m_{i}}^{i}}\|_{(i)}\\
&&& + \|y_{i}^{i, k_{m_{i}}^{i}} - y_{i}^{i, k_{m_{i}-1}^{i}}\|_{(i)}\\
&&& + \cdots + \|y_{i}^{i, k_{2}^{i}} - y_{i}^{i, k_{1}^{i}}\|_{(i)},\\
& < && m_{i} \varepsilon.
\end{aligned}
\]
Now take, as always, $\bar{x} = (\bar{y}_{1}^{1}, \ldots, \bar{y}_{n}^{n})$. Then we have, for all $i \in \{1, \ldots, n\}$,
\[
\begin{aligned}
\|f_{i}(\bar{x}) - f_{i}(\bar{y}^{i})\|_{(i)} & = &&&& \|f_{i}(\bar{y}_{1}^{1}, \ldots, \bar{y}_{n}^{n})\\
&&&&& - f_{i}(\bar{y}_{1}^{i}, \ldots, \bar{y}_{n}^{i})\|_{(i)},\\
& \le &&&& \|f_{i}(\bar{y}_{1}^{1}, \ldots, \bar{y}_{n}^{n})\\
&&&&& - f_{i}(\bar{y}_{1}^{i}, \bar{y}_{2}^{2}, \ldots, \bar{y}_{n}^{n})\|_{(i)}\\
&&& + && \|f_{i}(\bar{y}_{1}^{i}, \bar{y}_{2}^{2}, \ldots, \bar{y}_{n}^{n})\\
&&&&& - f_{i}(\bar{y}_{1}^{i}, \bar{y}_{2}^{i}, \bar{y}_{3}^{3}, \ldots, \bar{y}_{n}^{n})\|_{(i)}\\
&&& + && \cdots\\
&&& + && \|f_{i}(\bar{y}_{1}^{i}, \ldots, \bar{y}_{n-1}^{i}, \bar{y}_{n}^{n})\\
&&&&& - f_{i}(\bar{y}_{1}^{i}, \ldots, \bar{y}_{n}^{i})\|_{(i)}.\\
\end{aligned}
\]
Accounting Assumption~\ref{ass:var_coef} on $f$, it follows
\[
\|f_{i}(\bar{x}) - f_{i}(\bar{y}^{i})\|_{(i)} < \sum_{\substack{j = 1 \\ j \ne i}}^{n} \delta_{i,j}(f) m_{j} \varepsilon.
\]
Consider finally $\bar{y} = (\bar{y}^{1}, \ldots, \bar{y}^{n})$. Then we have
\[
\begin{aligned}
\|f(\bar{x}) - \bar{x}\|_{p} & = && \|f(\bar{x}) - g(\bar{y}) + g(\bar{y}) - \bar{x}\|_{p},\\
& \le && \|g(\bar{y}) - \bar{x}\|_{p}\\
&&& + \left( \sum_{i = 1}^{n} {\|f_{i}(\bar{x}) - f_{i}(\bar{y}^{i})\|_{(i)}}^{p} \right)^{1/p},\\
& < && \|g(\bar{y}) - \bar{x}\|_{p}\\
&&& + \left( \sum_{i = 1}^{n} \left( \sum_{\substack{j = 1 \\ j \ne i}}^{n} \delta_{i,j}(f) m_{j} \varepsilon \right)^{p} \right)^{1/p},\\
& < && \|g(\bar{y}) - \bar{x}\|_{p} + n^{1/p} \max_{i = 1}^{n} \sum_{\substack{j = 1 \\ j \ne i}}^{n} \delta_{i,j}(f) m_{j} \varepsilon.
\end{aligned}
\]
Applying \eqref{eq:max_m}, which means $m_{j} \le \eta$, and using Notation~\ref{not:max_var_coef}, we conclusively obtain
\[
\|f(\bar{x}) - \bar{x}\|_{p} < \|g(\bar{y}) - \bar{x}\|_{p} + n^{1/p} \eta \delta(f) \varepsilon.
\]
\end{proof}

Relatively to weighted maximum norms, defined on $X$ by
\[
\|x\|_{\infty}^{w} = \max_{i = 1}^{n} \frac{\|x_{i}\|_{(i)}}{w_{i}}, \quad w \in (\mathbb{R}^{+*})^{n},
\]
let us assume that $f$ is contractive, i.e.:
\begin{assumption}
\label{ass:wmn}
There exists a real $\alpha < 1$ such that
\[
\|f(x) - f(x')\|_{\infty}^{w} \le \alpha \|x - x'\|_{\infty}^{w}, \quad \forall x, x' \in X.
\]
\end{assumption}
Then, one may want to apply the following practical result.
\begin{proposition}
\label{prop:nf_ais4_wmn}
Let a sequence $\{x^{k}\}_{k \in \mathbb{N}}$, satisfying the asynchronous iterations model \eqref{eq:ai}, be generated by a computation $\mathcal{C}$ involving $n$ processes. Let sequences $\{y^{1,k}\}_{k \in \mathbb{N}}, \ldots, \{y^{n,k}\}_{k \in \mathbb{N}}$ be defined by the rewriting \eqref{eq:ai2}. Let, at last, $\bar{y}^{1}, \ldots, \bar{y}^{n}$ be the vectors returned by an execution of the non-FIFO AIS protocol 4 on $\mathcal{C}$. Then, under Assumptions~\ref{ass:ais2}, \ref{ass:sf}, \ref{ass:nf_ais} and~\ref{ass:wmn} :
\[
\|g(\bar{y}) - \bar{x}\|_{\infty}^{w} \le  \varepsilon = \frac{\varepsilon'}{1 + \eta \min_{i=1}^{n} w_{i}}
\]
implies
\[
\|f(\bar{x}) - \bar{x}\|_{\infty}^{w} < \varepsilon',
\]
with $\bar{y} = (\bar{y}^{1}, \ldots, \bar{y}^{n})$, $\bar{x} = (\bar{y}_{1}^{1}, \ldots, \bar{y}_{n}^{n})$ and $\varepsilon' \in \mathbb{R}$.
\end{proposition}
\begin{proof}
Considering the proof of Proposition~\ref{prop:nf_ais4}, we recall
\[
\|\bar{y}_{i}^{i} - \bar{y}_{i}^{j}\|_{(i)} < m_{i} \varepsilon, \quad i, j \in \{1, \ldots, n\}.
\]
According to Assumption~\ref{ass:wmn}, we have
\[
\|f(\bar{x}) - f(\bar{y}^{i})\|_{\infty}^{w} \le \alpha \|\bar{x} - \bar{y}^{i}\|_{\infty}^{w}, \quad \forall i \in \{1, \ldots, n\},
\]
and then, in particular,
\[
\begin{aligned}
\|f_{i}(\bar{x}) - f_{i}(\bar{y}^{i})\|_{(i)} & \le w_{i}~\alpha \|\bar{x} - \bar{y}^{i}\|_{\infty}^{w},\\
& \le w_{i}~\alpha \max_{j = 1}^{n} \frac{\| \bar{y}_{j}^{j} - \bar{y}_{j}^{i} \|_{(i)}}{w_{j}},\\
& < w_{i}~\alpha \max_{j = 1}^{n} \frac{m_{j}}{w_{j}} \varepsilon.
\end{aligned}
\]
It follows
\[
\begin{aligned}
\|f(\bar{x}) - \bar{x}\|_{\infty}^{w} & \le \|g(\bar{y}) - \bar{x}\|_{\infty}^{w} + \max_{i = 1}^{n} \frac{\|f_{i}(\bar{x}) - f_{i}(\bar{y}^{i})\|_{(i)}}{w_{i}},\\
& < \|g(\bar{y}) - \bar{x}\|_{\infty}^{w} + \alpha \max_{j = 1}^{n} \frac{m_{j}}{w_{j}} \varepsilon.
\end{aligned}
\]
Accounting $m_{j} \le \eta$ and $\alpha < 1$, we deduce
\[
\|f(\bar{x}) - \bar{x}\|_{\infty}^{w} < \|g(\bar{y}) - \bar{x}\|_{\infty}^{w} + \eta \max_{i = 1}^{n} \frac{1}{w_{i}} \varepsilon.
\]
Then, by ensuring $\|g(\bar{y}) - \bar{x}\|_{\infty}^{w} \le \varepsilon$, and taking
\[
\varepsilon = \frac{\varepsilon'}{1 + \eta \min_{i=1}^{n} w_{i}},
\]
we conclusively satisfy
\[
\begin{aligned}
\|f(\bar{x}) - \bar{x}\|_{\infty}^{w} & < \varepsilon + \eta \min_{i = 1}^{n} w_{i} \varepsilon,\\
& < \varepsilon'.
\end{aligned}
\]
\end{proof}

\section{Numerical results}
\label{sec:nr}

\subsection{Problem and experimental settings}

\begin{figure}[!t]
\centering
\includegraphics[scale=0.2]{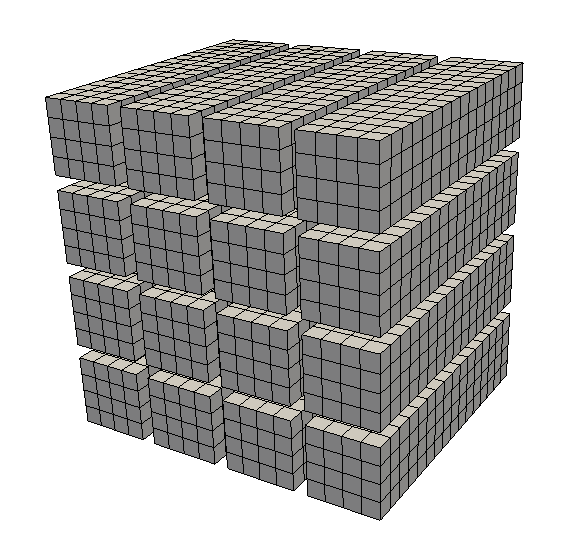}
\caption{Domain discretization and partitioning (16 sub-domains).}
\label{fig:partition}
\end{figure}

We are now interested in showing some experimental behavior of such asynchronous iterations snapshot protocols. For that, we consider the convection-diffusion problem
\[
\frac{\partial u}{\partial t} - \nu \Delta u + \vec{a}.\nabla u = s, \quad t \in \mathbb{R}^{+},
\]
where $u$ and $s$ are functions defined on $\mathbb{R}^{+} \times ([0, 1])^{3}$. Conditions and parameters are set to arbitrary values
\[
\left\{
\begin{array}{lll}
u(0,x,y,z) & = & 0, \quad \forall x, y, z \in (0, 1),\\
u(t,x,y,z) & = & 0, \quad \forall x, y, z \in \{0, 1\}, \forall t \in \mathbb{R}^{+},\\
\nu & = & 0.5,\\
\vec{a} & = & (0.1, -0.2, 0.3),
\end{array}
\right.
\]
just as the function $s$ given by
\[
s(t,x,y,z) = \sin(x) \sin(y) \sin(z),
\]
$\forall x, y, z \in [0, 1]$, $\forall t \in \mathbb{R}^{+}$. By using a finite-difference discretization and the backward Euler integration scheme, we obtain a sparse linear system
\[
\mathcal A U^{t_{i}} = B^{t_{i}, t_{i-1}},
\]
with $U^{t_{i}}, B^{t_{i}, t_{i-1}} \in \mathbb R^{m}$, $m \in \mathbb N$, at each time $t_{i} \in \mathbb{R}^{+}$, $i \in \mathbb{N}^{*}$, $t_{0} = 0$, for which we find an approximated solution $\widetilde U^{t_{i}}$ by means of successive relaxations of the form
\[
U^{t_{i}, k+1} := \mathcal M^{-1} \mathcal N U^{t_{i}, k} + \mathcal M^{-1} B^{t_{i}, t_{i-1}},
\]
with $k \in \mathbb{N}$, and $\mathcal A = \mathcal M - \mathcal N$ being a convergent splitting. While plenty of parallel executions were conducted using synchronous and asynchronous iterations $k$, we comment here only few of them which however accurately represent the overall results. Figure~\ref{fig:partition} illustrates the geometrical discretization and distribution of the domain $([0, 1])^{3}$ over parallel processes.
Each process handles exactly one sub-domain, and the number of processes always equals the number of processor cores used. Most of the simulations have been run for 5 time steps of size $\Delta t = 0.01$. We implemented the synchronous and asynchronous iterative methods using JACK~\cite{MagGBen2016}, our MPI-based communication library where we additionally introduced the various convergence detection methods.

\subsection{Effectiveness}

First experiments are led on a cluster of 68 nodes SGI Altix ICE 8400 LX with Quad Data Rate (QDR) Infiniband interconnect (40 Gbit/s). Each node consists of two 6-cores Intel Xeon X5650 Central Processing Units (CPU) at 2.66 GHz, and 21 GB Random Access Memory (RAM) allocated to parallel jobs. The Message Passing Interface (MPI) library SGI-MPT is loaded as communication middleware.

On practical aspects, we make few remarks about the proposed methods. First, non-FIFO AIS protocols 1 (NFAIS1) and 2 (NFAIS2) are very close to AIS protocols 1 (AIS1) and 2 (AIS2), respectively, and differ only on the content of the marker. Furthermore, AIS2 turns out to be a particular instance of the non-FIFO AIS protocol 5 (NFAIS5), when one consider $\eta = 0$. Second, the non-FIFO AIS protocol 4 (NFAIS4) is a generalization of NFAIS5, based on a behavior not likely to occur in most single-site high performance computing platforms. At last, the non-FIFO AIS protocol 3 (NFAIS3) is designed for very specific circumstances where markers exchange in NFAIS2 is to be avoided. Table~\ref{tab:residual} thus summarizes accuracy results of NFAIS1, NFAIS2 and NFAIS5, which are similar to the other AIS protocols.
\begin{table}[!t]
\caption{Effectiveness of AIS protocols, with residual threshold set to 1e-6.}
\label{tab:residual}
\centering
\begin{tabular}{|c|c|c|}
\hline
& Sync. iter. & NFAIS1\\
\begin{tabular}{cc}
$n$ & $\sqrt[3]{m}$\\
\hline
48 & 150\\
120 & 150\\
240 & 150\\
240 & 180\\
360 & 180\\
504 & 180\\
\end{tabular}
&
\begin{tabular}{cc}
$\min r_{i}$ & $\max r_{i}$\\
\hline
8.3e-7 & 8.3e-7\\
8.3e-7 & 8.3e-7\\
8.3e-7 & 8.3e-7\\
8.3e-7 & 8.3e-7\\
8.3e-7 & 8.3e-7\\
8.3e-7 & 8.3e-7\\
\end{tabular}
&
\begin{tabular}{cc}
$\min r_{i}$ & $\max r_{i}$\\
\hline
4.6e-7 & 6.9e-7\\
3.3e-7 & 5.0e-7\\
4.6e-7 & 5.6e-7\\
4.8e-7 & 6.5e-7\\
4.6e-7 & 5.5e-7\\
4.6e-7 & 5.8e-7\\
\end{tabular}\\
\hline
\end{tabular}
\hfill\\
\hfill\\
\hfill\\
\begin{tabular}{|c|c|c|}
\hline
& NFAIS2 & NFAIS5\\
\begin{tabular}{cc}
$n$ & $\sqrt[3]{m}$\\
\hline
48 & 150\\
120 & 150\\
240 & 150\\
240 & 180\\
360 & 180\\
504 & 180\\
\end{tabular}
&
\begin{tabular}{cc}
$\min r_{i}$ & $\max r_{i}$\\
\hline
5.4e-7 & 6.7e-7\\
4.6e-7 & 6.1e-7\\
3.8e-7 & 6.3e-7\\
4.5e-7 & 5.6e-7\\
5.0e-7 & 5.6e-7\\
4.8e-7 & 5.5e-7\\
\end{tabular}
&
\begin{tabular}{cc}
$\min r_{i}$ & $\max r_{i}$\\
\hline
5.2e-7 & 6.1e-7\\
5.2e-7 & 6.5e-7\\
4.8e-7 & 6.2e-7\\
4.7e-7 & 7.2e-7\\
4.3e-7 & 6.4e-7\\
5.5e-7 & 5.9e-7\\
\end{tabular}\\
\hline
\end{tabular}
\hfill\\
\hfill\\
\hfill\\
$r_{i} = \| \mathcal A \widetilde U^{t_{i}} - B^{t_{i}, t_{i-1}} \|_{\infty}$, $\quad \widetilde U^{t_{i}}, B^{t_{i}, t_{i-1}} \in \mathbb R^{m}$.\\
$n$ : number of processors.
\end{table}
Not surprisingly, as shown by these test cases, we did not face premature termination for any of our simulation runs. It is even noticeable that for any of the featured termination methods, the final residual tends to revolve around 5.5e-7 ($\pm$ 1e-7), regardless of both the number of processor cores and the size of the linear system. Such an experimental behavior strengthen the reliability of our protocols. Yet, compared to synchronous iterations which terminate at 8.3e-7, a few delay of 2.8e-7 ($\pm$ 1e-7) is introduced, however, as we shall see in the sequel, this does not prevent asynchronous iterations from terminating earlier than synchronous ones, in terms of execution time.

\subsection{Efficiency}

Table~\ref{tab:time_igloo} features total execution times and some mean measurements for one time step resolution. We introduce implementation of two other termination methods from \cite{Savari199639} (SB96) and \cite{BahiEtAl2008} (BCVC08), respectively, as described in Section~\ref{sec:rw}.
\begin{table}[!t]
\caption{Efficiency of AIS protocols, for 5 time step resolutions, $180^{3}$ unknowns and 504 cores.}
\label{tab:time_igloo}
\centering
\begin{tabular}{|c|c|c|c|c|}
\hline
method & time & mean \#it. & mean \#ss & mean $r_{i}$\\
\hline
\begin{tabular}{c}
Sync. iter.\\
SB96 \cite{Savari199639}\\
BCVC08 \cite{BahiEtAl2008}\\
NFAIS1\\
NFAIS2\\
NFAIS5
\end{tabular}
&
\begin{tabular}{r}
806\\
651\\
642\\
626\\
620\\
624
\end{tabular}
&
\begin{tabular}{r}
131867\\
182004\\
180782\\
176203\\
174429\\
175072
\end{tabular}
&
\begin{tabular}{r}
131867\\
13\\
8\\
1143\\
107\\
111
\end{tabular}
&
\begin{tabular}{r}
8.3e-7\\
8.5e-7\\
5.3e-7\\
5.2e-7\\
5.1e-7\\
5.6e-7
\end{tabular}\\
\hline
\end{tabular}
\hfill\\
\hfill\\
\hfill\\
time : total execution time, in seconds.\\
\#it. : maximum number of local iterations over the set of processes.\\
\#ss. : number of snapshots.
\end{table}
While discussing the effectiveness of these other methods is beyond the scope of this paper, we successfully verify that our AIS protocols do not introduce larger termination delays, regarding both execution times and maximum numbers of iterations. The maximum number of iterations over the set of processes quite well describes the resolution speed, as it produces the same ranking than execution time.

It is noticeable that our approach was more efficient despite a higher number of snapshots. Indeed, as the communication overhead cost is very low and that our methods run faster (only one reduction operation), they are more often executed to more quickly detect the actual convergence time, without impacting the iterations speed.

A part of the experiments involving much more processor cores has been conducted on another cluster of 5040 nodes Bullx B510, also with QDR Infiniband interconnect. Each node consists of two 8-cores Intel Sandy Bridge E5-2680 CPUs at 2.7 GHz, and 64 GB RAM. The Bullxmpi (OpenMPI) library is used as communication middleware. Here, we present in Table~\ref{tab:time_curie} some results for AIS1 protocol in an environment which however does not surely satisfy the FIFO assumption.
\begin{table}[!t]
\caption{Efficiency of AIS protocols, for 5 time step resolutions and more than 1000 cores.}
\label{tab:time_curie}
\centering
\begin{tabular}{|c|c|c|}
\hline
& Sync. iter. & SB96\\
\begin{tabular}{cc}
$n$ & $\sqrt[3]{m}$\\
\hline
1024 & 180\\
2048 & 185\\
5600 & 185
\end{tabular}
&
\begin{tabular}{rc}
time & mean $r_{i}$\\
\hline
251 & 8.3e-7\\
453 & 8.3e-7\\
530 & 8.4e-7
\end{tabular}
&
\begin{tabular}{rc}
time & mean $r_{i}$\\
\hline
132 & 7.0e-7\\
195 & 7.7e-7\\
112 & 2.9e-7
\end{tabular}\\
\hline
\end{tabular}
\hfill\\
\hfill\\
\hfill\\
\begin{tabular}{|c|c|c|}
\hline
& BCVC08 & AIS1\\
\begin{tabular}{cc}
$n$ & $\sqrt[3]{m}$\\
\hline
1024 & 180\\
2048 & 185\\
5600 & 185
\end{tabular}
&
\begin{tabular}{rc}
time & mean $r_{i}$\\
\hline
126 & 8.8e-7\\
185 & 7.4e-7\\
108 & 5.3e-7
\end{tabular}
&
\begin{tabular}{rc}
time & mean $r_{i}$\\
\hline
124 & 7.0e-7\\
179 & 8.5e-7\\
99 & 8.1e-7
\end{tabular}\\
\hline
\end{tabular}
\end{table}
First, we see that, with such a data transfer rate, Assumption~\ref{ass:sf} could be considered for AIS1 as well, with $\eta$ sufficiently small to avoid premature termination, even without implementing some adaptation of Proposition~\ref{prop:nf_ais4_wmn}. Secondly, it turned out that this slightly weakened version of AIS1 led to final residuals much closer to synchronous iterations ones, compared to results in Table~\ref{tab:residual}. At last, regarding execution times, its efficiency is confirmed, again compared to existing methods.

\section{Conclusion}
\label{sec:conclu}

Asynchronous iterations raise a non-trivial convergence detection issue that has been tackled in many various ways. Very few existing termination protocols are based on the computation of a global residual error, while mostly, more or less robust heuristics have been investigated. The most prominent approaches however require to perform two reduction operations, while we managed here to achieve effective convergence detection, using only one.
On practical aspects, it is noticeable that highly robust heuristics not based on global residual lead to quite intrusive, and often complicated, solutions which do not necessarily provide a substantial efficiency gain.

We proposed in this paper seven new asynchronous iterations termination methods based on global residual, under various communication models.
For FIFO communication environments, we proposed two protocols, AIS1 and AIS2, which we extended as NFAIS1 and NFAIS2 to any arbitrary non-FIFO communication model.
Rightly considering that FIFO communication is however essential for computation messages in parallel iterative processes, we exhibited a possible fifth protocol (NFAIS3) which avoids control messages in a context where the FIFO delivering is not guaranteed for messages of different types. This solution can however be slightly intrusive at implementation, and should be considered if marker-based non-FIFO methods are not easily applicable.
We then characterized a general non-FIFO model where, on every channel (in one direction), the number of messages that a given message can cross is bounded. The arbitrary non-FIFO model actually corresponds to the particular case where this maximum number always exceeds the number of messages emitted. We showed here how strong heuristics (NFAIS4 and NFAIS5) could be used to avoid including computation data into control messages, which constitutes an improvement of NFAIS1 and NFAIS2, in terms of communication overhead costs. We formally established the reliability of these heuristics, providing a practical way of accurately setting the convergence residual threshold.
Finally, experiments on supercomputers confirmed the effectiveness and efficiency of our approach versus prominent existing methods.

\section*{Acknowledgments}

A part of this work was performed using HPC resources from GENCI-TGCC (Grant 2014-t2014069065).

\bibliography{ref}
\bibliographystyle{abbrv}

\end{document}